\documentclass[twocolumn,showpacs,preprintnumbers,amsmath,amssymb]{revtex4-1}

\usepackage{graphicx}
\usepackage{dcolumn}
\usepackage{bm}

\begin{document}

\preprint{}

\title{
Light nuclei under magnetic field and the lithium problem
}

\author{Nodoka~Yamanaka$^{1,2}$}
  \affiliation{$^1$Department of Physics, Graduate School of Science, Tohoku University, Sendai 980-8578, Japan}
  \affiliation{$^2$Nishina Center for Accelerator-Based Science, RIKEN, Wako 351-0198, Japan}
  \email{nodoka.yamanaka@riken.jp}

\date{\today}

\begin{abstract}
We analyze the effect of the magnetic field on the proton and neutron density distributions of the nuclei $^2$H, $^3$H, $^3$He, which are calculated ab initio, and the $^6$Li nucleus in the $\alpha$-cluster model.
It is found that the asymptotic exponential damp of the probability density at long distance is modified, and that the linear component of the exponent with respect to the magnetic field yields the leading contribution, while those of the second derivative and the confining magnetic force are subleading.
Due to the linear dependence of the exponent, fluctuating magnetic fields always enhance the tunneling rate, i.e. the cross section of low energy nuclear reactions which occur at large separation across the Coulomb barrier.
While this mechanism cannot suppress the production of $^7$Be in the early Universe, it has the potential to resolve the lithium problem either by increasing the reaction rate of $^7$Be + $p$ $\to$ $^8$B at the bigbang nucleosynthesis era if the magnetic field at this time was sufficiently strong, or by correcting the systematically enlarged $^4$He + $^3$He $\to$ $^7$Be cross section by the unwanted magnetic field generated in the nuclear experimental setup, which was so far used as the input of the simulation of bigbang nucleosynthesis.
\end{abstract}

\pacs{26.35.+c, 27.10.+h, 13.40.-f, 21.60.Gx}

\maketitle

\section{\label{sec:intro}Introduction}

The bigbang nucleosynthesis (BBN) is one of the most successful prediction of cosmology \cite{Peebles:1966rol,Peebles:1966zz,Wagoner:1966pv,Smith:1992yy,Schramm:1997vs,Coc:2003ce,Cyburt:2015mya,Bertulani:2016eru,Mathews:2017xht,Pitrou:2018cgg,Fields:2019pfx,Reichert:2023xqy}, but there are still important open problems.
While the abundances of the protons (75\%) and $^4$He (25\%) are well explained by theoretical calculations, the predicted amount of $^7$Li nuclei (ratio between protons $^7$Li/H = $5.1 \times 10^{-10}$) is not agreeing with the observed one ($^7$Li/H = $1.6 \times 10^{-10}$) \cite{Sbordone:2010zi}.
The quantity of $^7$Li is evaluated by counting them at the surface of metal-poor and  low temperature stars, where it is expected that the number of $^7$Li did not evolve much since the BBN due to low nuclear reaction rate \cite{Spite:1982dd}, although there are still debates in their evolution in those stars \cite{Richard:2004pj,Piau:2006sw,Fu,Fields:2022mpw}.

On the experimental side, much effort has been devoted to accurately measure the nuclear reactions contributing to the lithium problem \cite{NaraSingh:2004vj,Descouvemont:2004cw,Gyurky:2007qq,LUNA:2007ffz,Brown:2007sj,DiLeva:2009zz,Fields:2011zzb,Bordeanu:2013coa,Kawabata:2017zpa,nTOF:2018vnr,Szucs:2019vuv,Hayakawa:2021jxf,Toth:2023boi}.
There are also several widely discussed theoretical approaches to resolve the lithium problem.
The most popular one was to add long-lived particles which do not belong to the standard model of particle physics so as to disturb the nuclear reactions \cite{Ellis:1984eq,Lindley:1984bg,Kawasaki:1986my,Scherrer:1987rr,Ellis:1990nb,Hamaguchi:2007mp,Jedamzik:2007cp,Kusakabe:2007fv,Pospelov:2010cw,Pospelov:2010hj,Kawasaki:2010yh,Cyburt:2012kp,Coc:2013eha,Kusakabe:2014moa,Goudelis:2015wpa,Fradette:2017sdd,Kawasaki:2017bqm,Alcaniz:2019kah,Boyarsky:2020dzc,Yeh:2022heq,Chang:2024mvg,Akhmedov:2024fpw,Chen:2024cla,Akita:2024nam,Ganguly:2025mdi}.

Recently, the effect of the magnetic field on the BBN was also discussed.
It was found that the primordial magnetic field may reduce the $^7$Li yield by deviating the thermal velocity distribution \cite{Luo:2018nth}, thus providing a partial resolution to the lithium problem.
It was also argued that the Coulomb barrier between charged nuclei are effectively decreased by the magnetic effect on plasma electrons \cite{Park:2023tmf}.
The primordial magnetogenesis is still not settled, but many attractive mechanisms which could have occurred before the BBN were conceived, such as the phase transitions or the inflation \cite{Thorne:1967zz,Matese:1969zz,Turner:1987bw,Quashnock:1988vs,Vachaspati:1991nm,Ratra:1991bn,Cheng:1993kz,Cheng:1996yi,Sigl:1996dm,Grasso:2000wj,Maeda:2008dv,Demozzi:2009fu,Kandus:2010nw,Kahniashvili:2010wm,Yamazaki:2012pg,Kobayashi:2014sga,Ellis:2019tjf,Di:2020kbw,Olea-Romacho:2023rhh,Yanagihara:2023qvx,Abramson:2025wyh}.
Theoretical studies of the magnetized plasma also progressed recently.
Self-consistent kinetic simulations of the plasma are now possible, and the spontaneous generations of magnetic field via the Weibel instability, turbulence, and the dynamo effect have been derived \cite{Zhou:2023eqo}.
While it is claimed that the magnetic energy saturates at the percent level of the plasma kinetic energy after the Weibel instability \cite{Takamoto:2018zgk}, the full kinetic simulation of the ion-electron plasma is still not available, and some enhancement of the final magnetic field is also expected after the dynamo stage \cite{Grasso:2000wj,Dimopoulos:1996nq,Brandenburg:2004jv,Chirakkara:2024ehm}.
It is then highly probable that the Universe at the BBN era was filled of strong magnetic field with a strength of the order of the actual (squared) temperature or higher, which may change the structure of light nuclei.
Essential questions that we might therefore ask are how the nuclear structure is modified, whether this modification affects the nuclear reactions contributing to the BBN or not, and eventually whether this resolves the lithium problem or not.

The structure of quantum systems under magnetic field may be studied by solving the many-body Schr\"{o}dinger equation with suitable interactions.
In this work, we propose to solve the quantum mechanics of light nuclei with the constituent nucleons interacting with an external magnetic field and to analyze their structure.
This setup has already been applied to the heavy quark systems \cite{Suzuki:2016kcs,Yoshida:2016xgm}, where the Gaussian expansion method \cite{Hiyama:2003cu} has been used.
This method has been successful in describing many physical systems \cite{Kameyama:1989zz,Hiyama:2000jd,Kamada:2001tv,Hamaguchi:2007mp,Kusakabe:2007fv,Funaki:2008gb,Hiyama:2010zzd,Hiyama:2012sma,Ohtsubo:2013wiz,Kusakabe:2014moa,Hiyama:2014kia,Yamanaka:2015qfa,Yamanaka:2015ncb,Hiyama:2016nwn,Yamanaka:2016fjj,Yamanaka:2016umw,Suzuki:2016kcs,Yoshida:2016xgm,Cheng:2018txc,Hiyama:2018ukv,Brodsky:2018zdh,Yang:2019lsg,Hiyama:2019kpw,Yang:2020atz,Yamanaka:2020kjo,Liu:2021rtn,Ogawa:2024hkb,Wen:2025wit,Wu:2025bnm}, so we expect it to also work in the case of few-nucleon systems under magnetic field.
We therefore inspect light nuclei which contribute to the BBN up to the three bodies, i.e. the deuteron ($^2$H), $^3$H, $^3$He, and the $^6$Li nucleus, which may be handled as a three-body system if we assume the cluster model \cite{clusterreview1,clusterreview2,clusterreview3,clusterreview4,Horiuchi:2012laa,Funaki:2015uya}.

Since the magnetic field is an electromagnetic effect, the change of the structure should be small, especially if one considers it at the BBN era.
A general bound solution of the Schr\"{o}dinger equation with a potential which goes to a constant value at infinite distance damps exponentially, so we expect that the leading response against the external magnetic field is a linear shift of the exponent.
This small modification is not important within a few fm, but it may become relevant for low energy nuclear reactions which occur across the Coulomb barrier, separated by several tens of fm.
Therefore, even a minor shift in the exponent might enhance the cross section relative to the reference value without magnetic fields.
The purpose of this paper is to analyze such effects, and to suggest some resolutions to the lithium problem.

This paper is organized as follows.
In the next Section, we introduce the model setup, the Gaussian expansion method, and the calculated observables.
In Section \ref{sec:result}, we analyze the results obtained and discuss the implication to the lithium problem, including some potential resolutions.
In the final Section, we summarize the discussion.

\section{Setup of calculation}

\subsection{Model and interactions}

The Hamiltonian of a nonrelativistic nuclear system under external magnetic field $\vec B$ is given by
\begin{eqnarray}
H
&=&
\sum_i
\Biggl[
\frac{1}{2 m_N} \Bigl( \vec p_i - Q_i e \vec A \, \Bigr)^2 
- \vec \mu_i \cdot \vec B
\Biggr]
\nonumber\\
&&
+ 
\sum_{i<j} V ( \vec r_i - \vec r_j \, )
,
\label{eq:hamiltonianmagnetic}
\end{eqnarray}
where $\vec r_i$, $\vec p_i$, $Q_i e$, and $\vec \mu_i$ are the coordinate, the momentum, the electric charge, and the magnetic moment of the $i$-th nucleon, respectively.
The nucleon mass is $m_N = 939$ MeV.
The sums over $i$ and $j$ are taken up to the total nucleon number of the nucleus.
By choosing the gauge $\vec A (\vec r_i ) = \frac{1}{2} \vec B \times \vec r_i$ and ignoring the center of mass motion, we have
\begin{eqnarray}
H
&\approx &
\sum_i
\Biggl[
\frac{\vec p_i^{\ 2}}{m_N}
-  \Bigl( 
\frac{\mu_p + \mu_n}{2} \vec \sigma_i 
+\frac{\mu_p - \mu_n}{2} \vec \sigma_i \tau_{zi}
\Bigr)
\cdot \vec B
\Biggr]
\nonumber\\
&&
+\sum_k^Z 
\Biggl[
- \frac{e}{2 m_N} \vec B \cdot \vec L_k 
+ \frac{e^2}{32 m_N} \Bigl( \vec B \times \vec r_k \, \Bigr)^2 
\Biggr]
\nonumber\\
&&
+ \sum_{i<j} V ( \vec r_i - \vec r_j \, )
,
\label{eq:hamiltonianapprox}
\end{eqnarray}
where the angular momentum of the $k$-th proton is defined by $\vec L_k \equiv \vec r_k \times \vec p_k$ ($1 \le k \le Z$).
The spin and (the $z$-component of) the isospin unit operators acting on the $i$-th nucleon are denoted by $\vec \sigma_i$ and $\tau_{zi}$, respectively.
The magnetic moments of the proton and the neutron are given by $\vec \mu_p \approx 2.79 \times \frac{ e }{2 m_N} \vec \sigma$ and $\vec \mu_n \approx -1.91 \times \frac{e }{2 m_N} \vec \sigma$, respectively.
In this work we use eV$^2$ ($\approx 5.1\times 10^{-3}$T = 51 G) as the unit for the magnetic field.
Here $V ( \vec r \, )$ is the nucleon-nucleon ($N-N$) potential.
For the deuteron ($^2$H), $^3$H, and $^3$He, we use the realistic nuclear force Argonne $v18$ \cite{Wiringa:1994wb}.
We do not consider the three-nucleon force \cite{Coon:1978gr,Carlson:1983kq,Pieper:2001ap} in this work.

The $^6$Li nucleus is modeled within the framework of the $\alpha +n+p$ three-body cluster model.
We use the Kanada-Kaneko potential for the $\alpha - N$ interaction \cite{Kanada:1979hhp} fitted so as to reproduce the scattering phase shifts at low energy.
For the $N-N$ subsystem of the $^6$Li nucleus, we employ the Argonne $v8'$ interaction \cite{Wiringa:1994wb}.
The Pauli exclusion principle relevant in subsystems with $\alpha$ is taken into account by using the orthogonality condition model (OCM) \cite{ocm1,Saito:1969zz,ocm2}.
This may practically be formulated by effectively applying a strong repulsion to the forbidden two-nucleon states, as \cite{Kukulin:1995tsx}
\begin{equation}
V_{\rm Pauli}
=
\lim_{\lambda \rightarrow \infty}\sum_{f} 
\lambda \, | \, f \, \rangle  \langle \, f \, |
.
\end{equation}
For the $\alpha - N$ subsystem, the only forbidden state is $f=0s$, and it is just the ground state wave function of the harmonic oscillator potential with the range 1.358 fm which reproduces the size of the $\alpha$ cluster.
We practically adopt a large value for the coupling $\lambda$, e.g. $\lambda = 10^4$, but a careful inspection of the stability of the output is needed as it may interfere with the small magnetic interaction (a similar interference could also be seen in the calculation of the electric dipole moment \cite{Yamanaka:2015qfa,Yamanaka:2016itb,Lee:2018flm,Yamanaka:2019vec}).

\subsection{Gaussian expansion method}

The Gaussian expansion method consists of solving the nonrelativistic Schr\"{o}dinger equation \cite{Hiyama:2003cu}
\begin{eqnarray}
( H - E ) \, \Psi_{JM,II_z}  = 0 ,
\label{eq:schr7}
\end{eqnarray}
using the variational principle.
The Hamiltonian to diagonalize is Eq. (\ref{eq:hamiltonianapprox}).

For the deuteron, the wave function is expanded as
\begin{eqnarray}
\Psi_{J=1,M}
&=&
\sum_{nls}
C_{nls}
\bigl[
\phi_{nlm}({\vec r} \, )
\otimes \chi ( s ) \bigr]_{J=1, M}
, \ \ 
\label{eq:deuteronwf}
\end{eqnarray}
where the radial component $\phi$ is expanded as 
\begin{eqnarray}
\phi_{nlm}({\vec r}\, )
&=&
{\cal N}_{n l}
r^l \, e^{-(r/r_n)^2}
Y_{lm}({\widehat {r}\, })  \;  ,
\end{eqnarray}
with the normalization factor ${\cal N}_{n l}$ and the geometric progression for the Gaussian range parameters 
\begin{eqnarray}
r_n
&=&
r_{\rm min} a^{n-1} \quad 
(n=1 - n_{\rm max}) \;
.
\end{eqnarray}
Due to the antisymmetrization constraint, the orbital angular momentum $l$ and the spin $s$ can only take the combinations $(l,s) = (0,1), (2,1)$ in the summation of Eq. (\ref{eq:deuteronwf}) under the restriction of the isospin $I=0$.

When interactions with a directional external field are present, we may also consider the Gaussian expansion method with the anisotropic cylindrical coordinates.
The basis function is then a double expansion:
\begin{equation}
\Phi_{n m} (d , z , \theta)
=
{\cal N}_{n m} e^{i l_z \theta} d^{| l_z |} e^{-\beta_n d^2 - \gamma_m z^2 }
,
\end{equation}
where $d \equiv \sqrt{x^2 +y^2}$, and $\theta$ is the angular coordinate in the $xy$-plane.

\begin{figure}[htb]
\begin{center}
\includegraphics[width=8.5cm]{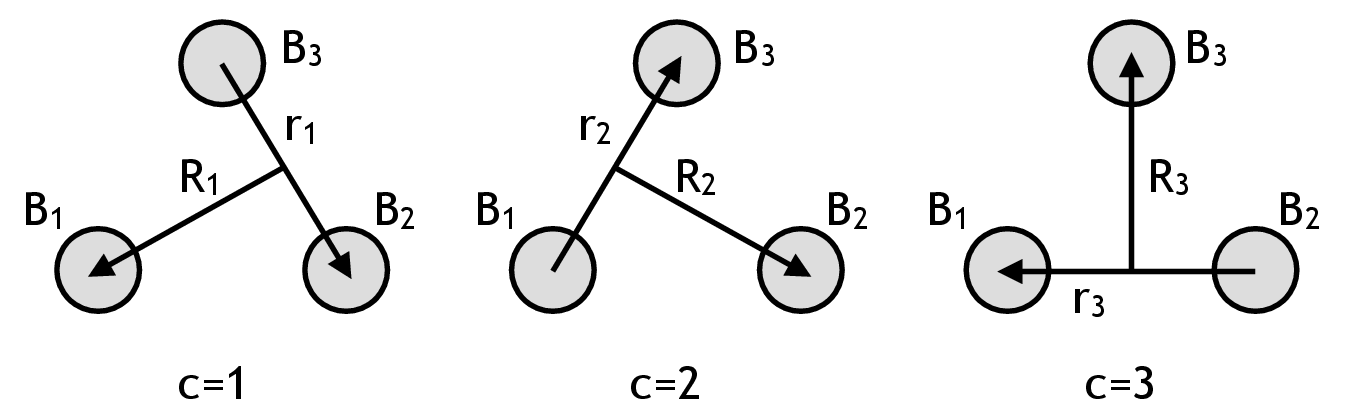}
\caption{\label{fig:jacobi}
The Jacobi coordinate of three-body systems.
}
\end{center}
\end{figure}

To calculate three-body systems, we need two radial coordinates to express the wave function.
For the identical 3-nucleon systems, the wave function is given as 
\begin{eqnarray}
&& \hspace{-3em}
\Psi_{JM, I, I_z}
\nonumber\\
&=&
\sum_{c=1}^3
\sum_{nl, NL}
\sum_{\Lambda} \sum_{T_c} \sum_{\Sigma} \sum_{s_c}
C^{(c)}_{nl,NL, \Lambda, s_c, \Sigma , T_c}
\nonumber\\
&& \hspace{1em}\times 
{\cal A} \biggl[
\Bigl[  \eta^{(c)} ( T_c ) \otimes \eta'^{(c)} ({\scriptstyle \frac{1}{2}} ) \Bigr]_{I={\scriptstyle \frac{1}{2}} ,I_z}
\nonumber  \\
&& \hspace{3.5em}
\times 
\Bigl[
[ \phi^{(c)}_{nl}({\vec r}_c) \otimes \psi^{(c)}_{NL}({\vec R}_c)]_\Lambda 
\nonumber\\
&& \hspace{5em}
\otimes \, \bigl[ \chi^{(c)} ( s_c ) \otimes \chi'^{(c)} ({\scriptstyle \frac{1}{2}} ) \bigr]_\Sigma \Bigr]_{J M}
\biggr]
, \ \ 
\label{eq:he7lwf}
\end{eqnarray}
where ($\eta$,$\eta'$), ($\phi$,$\psi$), and ($\chi$,$\chi'$) are the two isospin, radial, and spin components of the nucleon state according to the coupling of the Jacobi coordinate $c$ (see Fig. \ref{fig:jacobi}), respectively.
Here the operation of the antisymmetrizing operator $\cal{A}$ practically imposes the constraint of taking only odd $T+l+s$, under the condition that all the three Jacobi coordinates are summed, which means that the coefficients $C$ of Eq. (\ref{eq:he7lwf}) is independent of the channel $c$.
For the $^6$Li in the $\alpha$-cluster model, the particles are not all identical, so the antisymmetrization constraint only applies to the proton and neutron, and the coefficients $C$ are not all identical (if we take the particle $B_3$ to be the $\alpha$ cluster under the definition of Fig. \ref{fig:jacobi}, the channel $c=3$ has an independent coefficient while those of $c=1$ and $c=2$ are equal).

For the three-body case, we expand the two radial components $\phi$ and $\psi$ as 
\begin{eqnarray}
\phi_{nlm} ({\vec r}\, )
&=&
{\cal N}_{n l}
r^l \, e^{-(r/r_n)^2}
Y_{lm}({\widehat r})
,
\nonumber \\
\psi_{NLM} ({\vec R}\, )
&=&
{\cal N}_{NL}
R^L \, e^{-(R/R_N)^2}
Y_{LM}({\widehat {R}})
,
\end{eqnarray}
where the following geometric progressions for the Gaussian range parameters are used:
\begin{eqnarray}
r_n
&=&
r_{\rm min} a^{n-1} \ \ \ 
(n=1 - n_{\rm max})
,
\nonumber\\
R_N
&=&
R_{\rm min} A^{N-1} \ \ \ 
(N =1 - N_{\rm max})
.
\end{eqnarray}
For the three-body systems, there are infinitely possible orbital angular momentum channels, but the truncation is sufficient with $l, L, \Lambda \leq 2$ to obtain good convergence of the result.
We may also expand the three-body wave function in the cylindrical coordinate, but we do not introduce it here explicitly since it will not be used in this work.

\subsection{Nucleon density distribution of nuclei under magnetic field}

\begin{figure*}[htb]
\begin{center}
\includegraphics[width=8.5cm]{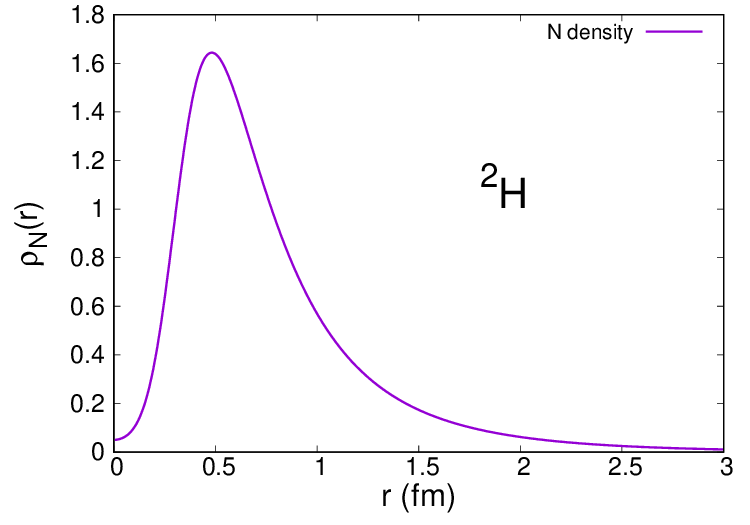}
\includegraphics[width=8.5cm]{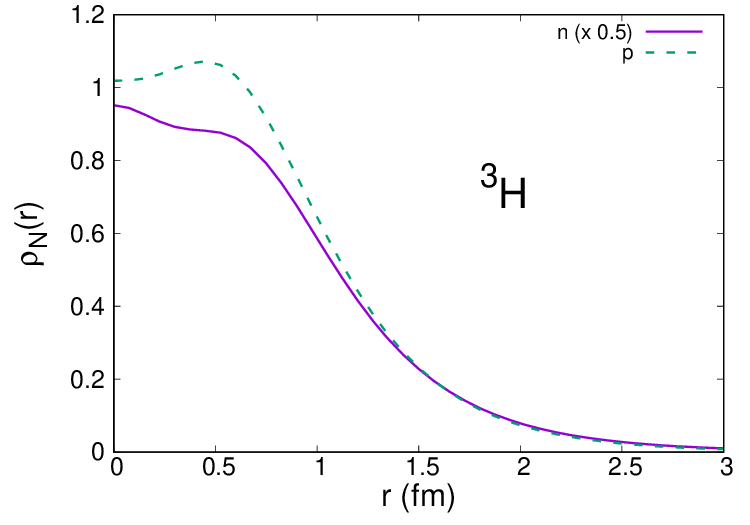}
\includegraphics[width=8.5cm]{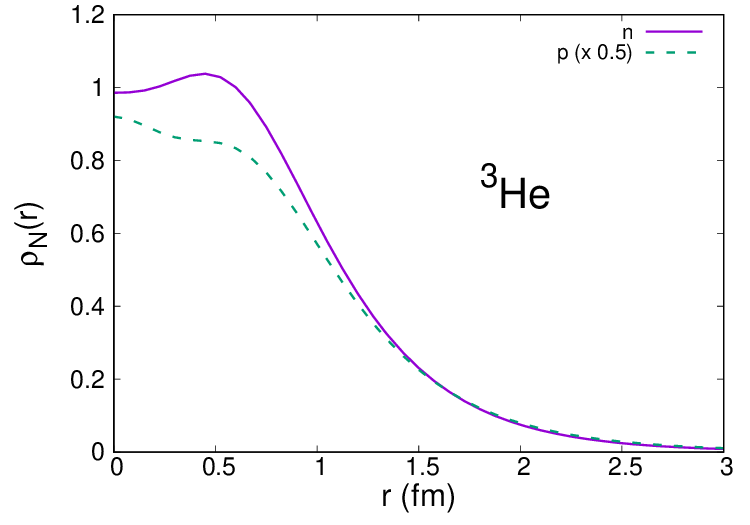}
\includegraphics[width=8.5cm]{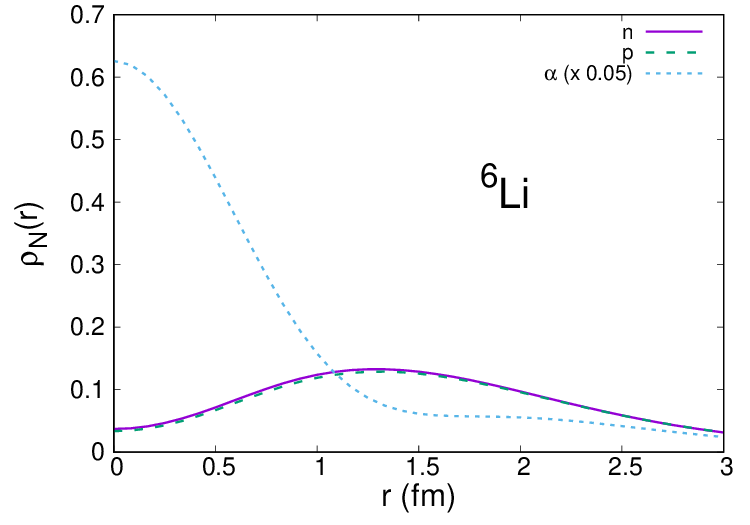}
\caption{\label{fig:density_distribution}
Density distributions of $^2$H, $^3$H, $^3$He, and $^6$Li.
The distance $r$ is the coordinate of the single nucleon with the center of mass as the origin.
}
\end{center}
\end{figure*}

In this work, we essentially calculate the one-nucleon density distribution of the nuclear system, given as
\begin{equation}
\rho_N (r)
=
\frac{1}{2 r^2}
\langle \Psi | \,
(1\pm \tau')
\delta [\, (A-1)R' /A - r \, ]
\, | \Psi \rangle
,
\label{eq:densitydefinition}
\end{equation}
where $R'$ is the last Jacobi coordinate (which relates the last nucleon to the center of mass of the remaining subsystem), $A$ the total nucleon number, and $\tau'$ is the isospin Pauli matrix which only acts to the isospin basis of the last nucleon $\eta'$ [see Eq. (\ref{eq:he7lwf})], with the positive (negative) sign for the proton $N=p$ (for the neutron $N=n$).
The density distribution $\rho_N (r)$ is then the probability density to find a nucleon $N$ at the distance $r$ from the center of mass, and we normalize it to one per nucleon.
We plot in Fig. \ref{fig:density_distribution} the short distance behaviors of the nucleon density distributions of $^2$H, $^3$H, $^3$He, and $^6$Li.
We note that for the $^6$Li nucleus, the nucleons inside the $\alpha$ cluster are not taken into account, but we plot instead the distribution of the $\alpha$ cluster.

To inspect the effect of the magnetic field, we calculate the nucleon density distributions of nuclei which are parallelly and antiparallelly polarized along it.
The energy of the nuclear system is then maximized or minimized due to the terms with $\vec B \cdot \vec \sigma$ and $\vec B \cdot \vec L$ of the Hamiltoninan (\ref{eq:hamiltonianapprox}) in the first order of $B$.
If we crudely model the nuclear potential by a square well with depth $V_0$, the wave function outside the nucleus behaves as $\sim e^{-r \sqrt{ 2 m_N |V_0| \pm (2 m_N \mu_N + Q_N e l_N ) B }}  $.
We therefore expect a modification of the exponential damp of the nucleon probability density, which might lead to a large enhancement or suppression if the nuclear reaction happens at long distance.
The real nuclear potential is of course not a square well and the exponent generally has logarithmic corrections, but the most relevant contribution is the exponential damp as seen above.

In the use of variational methods like the Gaussian expansion method, it is technically difficult to directly describe the long range tail of the wave function which has very small contribution to the energy of the system, and the behavior there is often unstable.
In this work, we fit by an exponential function the probability density at sufficiently large distance such that it is behaving exponentially, but still stable enough within the variational method.
We then Taylor expand the exponent by the external magnetic field up to the second order.
At this order, the magnetic confining interaction [term with $(\vec B \times \vec r \,)^2$ of Eq. (\ref{eq:hamiltonianapprox})] also contributes.
Its effect has to be calculated in the cylindrical basis.

\section{Results and discussions\label{sec:result}}

\begin{figure*}[htb]
\begin{center}
\includegraphics[width=8.5cm]{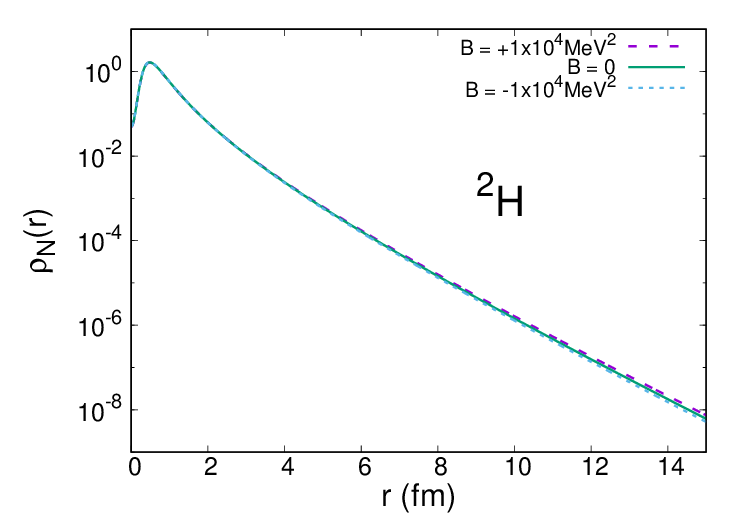}
\includegraphics[width=8.5cm]{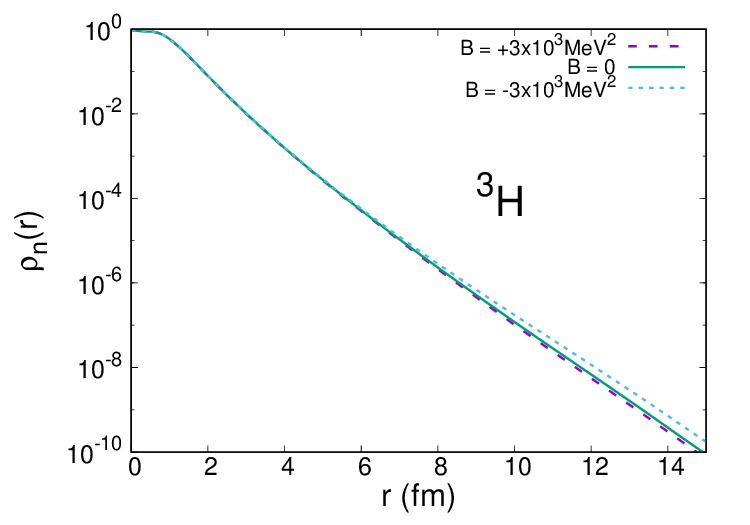}
\includegraphics[width=8.5cm]{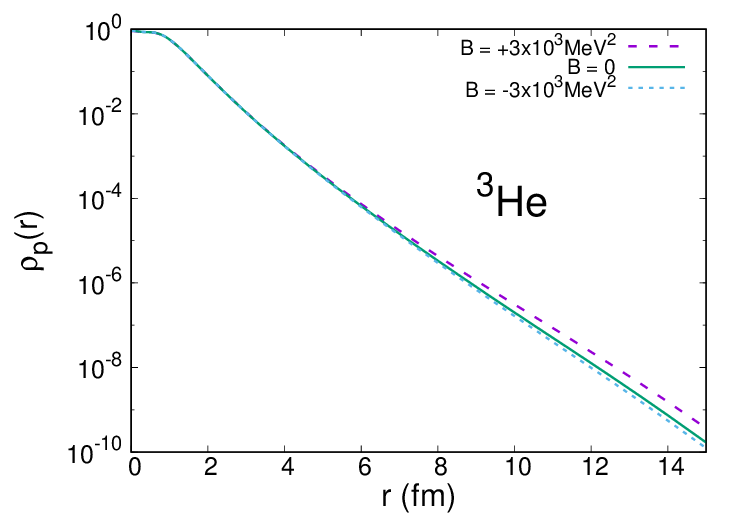}
\includegraphics[width=8.5cm]{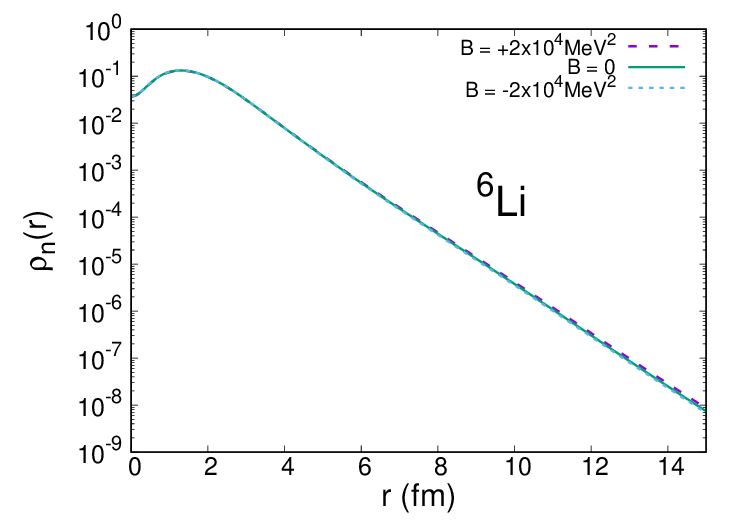}
\caption{\label{fig:triton_probability_large_magnetic_test}
The modifications of the neutron probability distributions for $^2$H, $^3$H, $^6$Li, and the proton one for $^3$He under large magnetic field (plotted in the logarithmic scale).
}
\end{center}
\end{figure*}

We now analyze the result of the calculation of the nucleon density distributions.
In Fig. \ref{fig:triton_probability_large_magnetic_test}, we plot their modification under large magnetic field.
We see deviations of the exponential damp at distance beyond $r= 10$ fm.
Since we are interested in the modification at large $r$, we define the following asymptotic relative factor (ARF)
\begin{equation}
\frac{\rho_N (r) |_{B_z=B} }{\rho_N (r) |_{B_z=0}}
\sim
e^{-C_N(B) r }
\ \ \ \ 
(r \to \infty)
,
\label{eq:ARF}
\end{equation}
where the magnetic field dependence of the exponent $C_N(B)$ is expanded up to the second order in this work, as $C_N(B) \approx C_N^{(1)} B + C_N^{(2)} B^2$.
We fit the one-nucleon probability density at two points, namely $r=7$ fm and $r=12$ fm, for which our calculation is sufficiently stable.
This relative factor has a practical meaning in the context of low energy nuclear reactions and the BBN.
By assuming that the tunneling rate is proportional to the nucleon probability density under the Coulomb barrier, the ARF (\ref{eq:ARF}) gives the enhancement/suppression of the nuclear fusion cross section as a function of the penetration and the magnetic field.

We first consider the effect of the linear terms (with $\vec B \cdot \vec \sigma$ and $\vec B \cdot \vec L$) of the Hamiltoninan (\ref{eq:hamiltonianapprox}).
We show the result of our fits for $^2$H, $^3$H, $^3$He, and $^6$Li for each nucleon in Table \ref{table:magneticdependence}.
We see that the neutron probability density of $^3$H and the proton one of $^3$He are particularly sensitive to the magnetic field.
On the other hand, the close coefficients of $^2$H and $^6$Li suggest that the $^6$Li nucleus, which is described as an $\alpha$+$^2$H cluster system, is also inheriting the sensitivity of $^2$H against the magnetic field.
We may also interpret the high sensitivity of $^3$H and $^3$He as compared to the deuteron by the compactness of the nuclear density.
Since compact systems have faster exponential falloff of the wave function at long distance, the shift due to the magnetic field, which is also an exponential effect, becomes relatively important.
This remark then suggests that compact nuclei have higher sensitivity against the magnetic field than nuclei having loosely bound subclusters, of course under the premise that they have open spin shell (the $^4$He nucleus is then uninteresting in the point of view of this discussion since it has a closed spin shell, even though it is the most compact nucleus).

Here we point out that the linear interaction of the magnetic field with the nucleon magnetic moment (interaction with $\vec B \cdot \vec \sigma$) does not contribute to the modification of the density distribution unless there is a mixing between the spin and the orbital angular momentum because otherwise all states will be shifted by a same constant energy which is just a change of the potential energy reference.
Our result then suggests that $^3$H and $^3$He have more configuration mixing of orbital angular momentum than $^2$H and $^6$Li.

In an environment with fluctuating magnetic field, the response is averaged.
It is important to remark that the net effect is {\it always enhanced}, since the exponent is linear in the magnetic field.
By assuming a fluctuation of $B_z = \pm B$, the averaged ARF becomes
\begin{equation}
\frac{ e^{-C_N^{(1)} B r} + e^{+C_N^{(1)} B r} }{2}  \ge 1
.
\label{eq:magneticenhance}
\end{equation}
This means that the fluctuation of the magnetic field enhances the cross section at the same order as a constant magnetic field background with the same amplitude.
We also make a caution that the coordinate $r$ used in this work is the distance from the center of mass [see Eq. (\ref{eq:densitydefinition})], so this might not exactly be the distance controlling the Coulomb barrier.

\begin{table}
\caption{
First and second order Taylor coefficients of the magnetic field dependent function $C_N(B)$.
For the deuteron ($^2$H), the proton and neutron probability densities are the same.
}
\begin{ruledtabular}
\begin{tabular}{l|c|l|l|}
Nucleus & $N$ & $C_N^{(1)}$ [MeV$^{-2}$.fm$^{-1}$] & $C_N^{(2)}$ [MeV$^{-4}$.fm$^{-1}$] \cr
\hline
$^2$H & $p,n$ & $1.5 \times 10^{-6}$ & $2.8 \times 10^{-12}$ \cr
\hline
$^3$H & $p$ & $2.3 \times 10^{-6}$ & $-1.7 \times 10^{-10}$ \cr
& $n$  & $-1.5 \times 10^{-5}$ & $2.6 \times 10^{-9}$ \cr
\hline
$^3$He & $p$ & $1.7 \times 10^{-5}$ & $2.3 \times 10^{-9}$ \cr
& $n$ & $1.7 \times 10^{-7}$ & $2.0 \times 10^{-10}$ \cr
\hline
$^6$Li & $p$ & $5.4 \times 10^{-7}$ & $5.0 \times 10^{-12}$ \cr
& $n$ & $6.1 \times 10^{-7}$ & $4.0 \times 10^{-12}$ \cr
\end{tabular}
\end{ruledtabular}

\label{table:magneticdependence}
\end{table}

\begin{figure}[htb]
\begin{center}
\includegraphics[width=8.5cm]{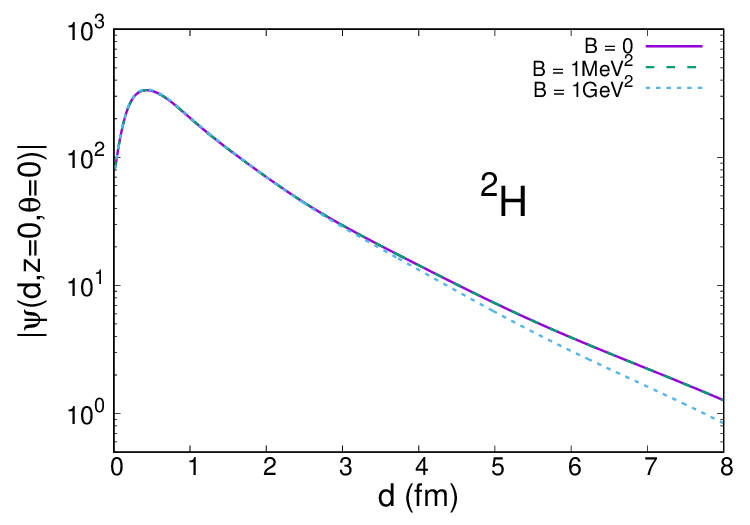}
\caption{\label{fig:deuteron_wavfct_cylindrical}
Modification of the wave function of $^2$H by the confining magnetic force in the cylindrical coordinate.
The curves for $B=0$ and $B=1$ MeV$^2$ are almost overlapping.
}
\end{center}
\end{figure}

We also see from Fig. \ref{fig:triton_probability_large_magnetic_test} that the slope of the exponential damp is not symmetric under the sign change of the magnetic field, which is due to the second order coefficient $C_N^{(2)}$.
This effect only becomes relevant under large magnetic field, and it is negligible in the context of the BBN.
We should also check the effect of the confining interaction [term with $(\vec B \times \vec r \,)^2$ of Eq. (\ref{eq:hamiltonianapprox})] which also contributes at the second order in $B$.
We display in Fig. \ref{fig:deuteron_wavfct_cylindrical} the result of the calculation of the wave function of the deuteron obtained using the cylindrical basis and the nuclear force of the modified Malfliet-Tjon I-III model \cite{Friar:1990zza,Hiyama:2003cu}. 
In this case, the probability at long distance is strictly suppressed since it is a confining force.
We also see that the modification starts to become relevant only by applying a very large magnetic field, of $O(1\,{\rm GeV}^2)$, so it is negligible in the context of the BBN as well.
We also note that the probability density of the $^4$He nucleus, which has a closed spin shell, is only modified by this mechanism, so that the effect of the magnetic field on $^4$He is negligible.

Now let us discuss the implication of our calculation to the BBN and to the lithium problem.
We remind that this is the yield of $^7$Be at the BBN era which has to be reduced in order to resolve this problem, since the $^7$Li nuclei existing in the late Universe (before the star formation) mostly originate from the decay of $^7$Be, and those generated at the BBN time ($\sim$ 1000 s after the birth of the Universe) are destroyed by the reaction $^7$Li + $p$ $\to$ $^4$He + $^4$He, so that they are actually subleading.
For this reason, the major approach to the lithium problem is to suppress $^4$He + $^3$He $\to$ $^7$Be, but the destruction of $^7$Be at the BBN era is also possible.
To solve the problem using the magnetic field $B$, we have to mainly consider collisions between charged particles, because $B$ only modifies the wave function at long distance where low energy reactions happen across the Coulomb barrier, while the neutron collisions are not disturbed.
As we saw in Eq. (\ref{eq:magneticenhance}), the nuclear fusion is always enhanced by the magnetic field, so we cannot suppress the yield of $^7$Be through the process $^4$He + $^3$He $\to$ $^7$Be.
The possible solution is then to enhance $^7$Be + $p$ $\to$ $^8$B (which finally becomes $\beta^+$ + $^4$He + $^4$He) at the BBN.
At this time, nuclei have a typical kinetic energy of 0.1 MeV, which corresponds to a penetration of 60 fm for the $^7$Be + $p$ collision.
It is known that $^7$Be is well described by the $\alpha$+$^3$He cluster structure, so its wave function is expected to have a similar $B$-dependence as that of the $^3$He nucleus, just like the resemblance between $^2$H and $^6$Li seen in this work (see Table \ref{table:magneticdependence}).
The $^7$Be wave function under magnetic field will have to be quantified in the future, ideally in an ab initio framework.
We also add that the neutron distribution of $^7$Li has a high sensitivity to the magnetic field according to our result (see Table \ref{table:magneticdependence}),  so it is possible that the destruction process $^7$Li + $p$ $\to$ $^4$He + $^4$He is also enhanced.
As we saw above, however, $^7$Li nuclei at the BBN era are much less numerous than $^7$Be, so this mechanism does not resolve the lithium problem.
To obtain a variation of the $^7$Be + $p$ reaction cross section by a factor two for the penetration of 60 fm, we need an external magnetic field $B\sim 800$\, MeV$^2$.
Although there are constraints on the magnetic field at the BBN era \cite{Thorne:1967zz,Matese:1969zz,Cheng:1993kz,Cheng:1996yi,Grasso:2000wj}, it is currently not exactly known, and mechanisms of enhancement such as the dynamo are still possible \cite{Zhou:2023eqo,Dimopoulos:1996nq,Chirakkara:2024ehm}.
There might be other relevant external factors which enhance the reaction rate such as the screening of the Coulomb barrier by the plasma electron, as argued in Ref. \cite{Park:2023tmf}.
The quantifications of the above points are left for future works.

An alternative nontrivial mechanism which might be the origin of the lithium problem is the systematic enhancement of the $^4$He + $^3$He $\to$ $^7$Be reaction cross section by the magnetic field generated in nuclear collision experiments.
In this case, the input data used in the calculation of the BBN were systematically increased, which eventually led to the large BBN yield prediction of $^7$Be.
The nuclear reactions were measured in fixed-target setup \cite{Parker:1963zz,Nagatani:1969mpa,Osborne:1982zz,Krawinkel,Robertson:1983zz,Alexander:1984gfc,Hilgemeier,NaraSingh:2004vj,Gyurky:2007qq,LUNA:2007ffz,Brown:2007sj,DiLeva:2009zz,Bordeanu:2013coa,Szucs:2019vuv,Toth:2023boi}, where the projectile passed close to many nuclei and electrons of the gas target or the window before hitting the target nucleus, so a large magnetic fluctuation might have been generated even though the average is almost null (as a similar setup, we have the heavy ion collision where the generation of strong magnetic field is predicted \cite{Rafelski:1975rf,Bzdak:2011yy,Deng:2012pc}).
As we emphasized in this work, the fluctuation enhances the exponential tail of the probability distribution at long distance.
Whether a fluctuation of $\delta B\sim 800$\, MeV$^2$ is generated in the experimental setup or not is a problem of the dynamics, and this needs to be quantified in a future work as well.

\section{Conclusion}

In this work, we calculated the modification of the probability densities of $^2$H, $^3$H, $^3$He, and $^6$Li by the magnetic field at long distance.
The two- and three-nucleon systems were described ab initio using the Argonne v18 potential, while the $^6$Li nucleus was handled in the $\alpha$ cluster model as a three-body system.
The wave function was calculated using the variational Gaussian expansion method.
Since it is difficult in variational methods to directly quantify the nuclear density at the penetration radius with the typical kinetic energy at the BBN, we fitted the exponent of the asymptotic exponential damp and Taylor expanded it up to the second order of the external magnetic field.
It was found that the linear coefficient is large for the proton density in $^3$He and for the neutron one in $^3$H.
We also checked that the second order term as well as the magnetic confining force have negligible contribution for small magnetic field, at least in the context of the BBN.

To resolve the lithium problem we proposed two solutions, both by enhancing nuclear reactions with the magnetic field.
The first solution is to increase the $^7$Be + $p$ $\to$ $^8$B $\to$ $^4$He + $^4$He + $\beta^+$ reaction rate at the BBN era.
The second solution consists of revising the experimental measurement of the $^4$He + $^3$He $\to$ $^7$Be collision which might have been affected by the magnetic fluctuation generated by the projectile passing close to non-reacting nuclei and electrons.
These two points have to be quantified in the future.

The change of the structure at long distance by the magnetic field and its fluctuation might also affect the experimental extractions of other nuclear scattering cross sections at low energy as well as the nucleosynthesis in supernovae or in neutron star mergers.
The inspection performed in this work then provides a potentially significant caution to the low energy physics of Coulomb barrier.

\begin{acknowledgments}
This research was conducted using the FUJITSU Supercomputer PRIMEHPC FX1000 and FUJITSU Server PRIMERGY GX2570 (Wisteria/BDEC-01) at the Information Technology Center, The University of Tokyo.
This work (project) was supported by the RIKEN TRIP initiative (Nuclear transmutation), and also partially by ERATO project-JPMJER2304.
\end{acknowledgments}

\twocolumngrid

\end{document}